\documentclass[10pt, journal, compsoc]{IEEEtran}
\usepackage{amsmath,amsfonts}
\usepackage{algorithmic}
\usepackage{algorithm}
\usepackage{array}
\usepackage[caption=false,font=normalsize,labelfont=sf,textfont=sf]{subfig}
\usepackage{textcomp}
\usepackage{stfloats}
\usepackage{url}
\usepackage{verbatim}
\usepackage{microtype}
\usepackage{graphicx}
\usepackage{cite}
\usepackage{siunitx}

\usepackage{xspace}
\usepackage[usenames,dvipsnames]{xcolor} %

\newcommand{\quotes}[1]{``#1''} %

\newcommand{\revision}[1]{\textcolor{black}{#1}}
\newcommand{\minorrevision}[1]{\textcolor{black}{#1}}

\newcommand{\DT} {\noindent{decision task}\xspace} %
\newcommand{\JT} {\noindent{judgment task}\xspace} %

\newcommand{\DA} {\noindent{decision accuracy}\xspace} %
\newcommand{\JA} {\noindent{judgment accuracy}\xspace} %

\newcommand{\condScenSports} 		{\noindent{sports}\xspace}
\newcommand{\condScenHumanitarian} 		{\noindent{humanitarian}\xspace}
    
\newcommand{\condVisQDP}        {\noindent{quantile dotplot}\xspace}
\newcommand{\condVisDensity} 	{\noindent{density plot}\xspace}
\newcommand{\condVisProbBar} 	{\noindent{probability bar}\xspace}

\begin{document}

\title{Decoupling Judgment and Decision Making:\\A Tale of Two Tails}

\author{Başak Oral, %
        Pierre~Dragicevic, %
        Alexandru Telea, 
        and~Evanthia Dimara%
\IEEEcompsocitemizethanks{\IEEEcompsocthanksitem B. Oral was with Utrecht University.\protect\\
E-mail: e.oral@uu.nl
\IEEEcompsocthanksitem P. Dragicevic was with  Inria Bordeaux.
 \IEEEcompsocthanksitem A. Telea, and E. Dimara were with Utrecht University.}

\thanks{Manuscript received January 10, 2023; revised September 29, 2023.}}

\markboth{Journal of \LaTeX\ Class Files,~Vol.~X, No.~X, mon~2023}%
{Shell \MakeLowercase{\textit{et al.}}: A Sample Article Using IEEEtran.cls for IEEE Journals}

\IEEEpubid{0000--0000/00\$00.00~\copyright~2023 IEEE}

\IEEEtitleabstractindextext{%
\begin{abstract}

Is it true that if citizens understand hurricane probabilities, they will make more rational decisions for evacuation? Finding answers to such questions is not straightforward in the literature because the terms \quotes{\emph{judgment}} and \quotes{\emph{decision making}} are often used interchangeably. This terminology conflation leads to a lack of clarity on whether people make suboptimal decisions because of inaccurate judgments of information conveyed in visualizations or because they use alternative yet currently unknown heuristics. To decouple judgment from decision making, we review relevant concepts from the literature and present two preregistered experiments (N=601) to investigate if the task (judgment \emph{vs.} decision making), the scenario (sports \emph{vs.} humanitarian), and the visualization (quantile dotplots, density plots, probability bars) affect accuracy. \revision{While experiment 1 was inconclusive, we found evidence for a difference in experiment 2. Contrary to our expectations and previous research, which found decisions less accurate than their direct-equivalent judgments, our results pointed in the opposite direction. Our findings further revealed that decisions were less vulnerable to status-quo bias, suggesting decision makers may disfavor responses associated with inaction. We also found that both scenario and visualization types can influence people's judgments and decisions. Although effect sizes are not large and results should be interpreted carefully, we conclude that judgments cannot be safely used as proxy tasks for decision making, and discuss implications for visualization research and beyond. Materials and preregistrations are available at \url{https://osf.io/ufzp5/?view_only=adc0f78a23804c31bf7fdd9385cb264f}.}
\end{abstract}

}

\maketitle

\section{Introduction}
\label{sec:introduction}

Imagine a user browsing a weather app on their phone that shows forecast data. They ask themselves two questions: \emph{i)} How likely is it that there will be heavy rainfall this afternoon? \emph{ii)} Should I carry an umbrella on my way to the doctor’s appointment? These questions seem related, so it is fair to assume that our \emph{judgment} of the forecast probability should directly influence our \emph{decision} to carry an umbrella. Yet, even if we assume that everyone can correctly derive the probability estimates from the visualization, other factors, such as our bag size, the umbrella weight, or our tolerance to mild rainfall, may influence our decision. More life-threatening decisions, such as if to evacuate a city based on the forecast of a hurricane strike or wildfire, are more arduous and suffer from cognitive biases\,\cite{Padilla2018}. Stressful situations can further introduce time pressure leading to inaccurate assessment of the visualized data\,\cite{Cheong2016}.

It is rather unclear if visualization users make suboptimal decisions because they perform inaccurate judgments of probabilities, or they use alternative yet currently unknown heuristics. More research is needed to explain \emph{why} people make seemingly suboptimal decisions. One barrier to answering this question is that researchers tend to use the terms judgment and decision interchangeably. For example, in a visualization study, participants were asked to decide, as Red Cross managers, on issuing blankets to help alpacas survive harsh weather conditions\,\cite{Castro}. Although the task was a decision, the interpretation of task accuracy was associated with the ability of participants to judge the probability of temperatures falling below \SI{32}{\degree\,\text{F}}. Other experiments, while aiming to focus on studying human decisions, exposed participants to visual judgments, such as finding the maximum average bar length\,\cite{Jardine2020} or finding a growth trend\,\cite{Kale2019}. The notions of judgment and decision making are often conflated in a single task which sometimes measures people’s decision performance and other times their ability to judge numerical estimates. More importantly, the initial goal of the study is not always in line with the underlying task, which makes the results hard to interpret.

To some extent, this conflated use of judgments and decisions is justifiable --  a judgment is a fact-based question that is easier to test than a decision. Yet, recent research gives preliminary evidence that people may need to use different visualizations for judgment and decision tasks. Kale \emph{et al.}\,\cite{Kale2021} presented participants with four different visualizations (quantile dotplots, density plots, HOPs, intervals), in a fantasy sports game. The task was to judge how probably a participant's team would score with or without a new player, as well as to decide whether to pay for the new player or not. They found that, while participants elicited the best decisions with densities and intervals, they judged better with quantile dotplots. Although the goal of the paper was not to compare judgments with decisions, the fact that the best visualization for judgment did not necessarily lead to better decisions reveals the need to decouple these two tasks. So, to clarify the confusion of judgment and decision making, it is important to separate judgment and decision tasks and to investigate to what extent they relate to each other. 

In this paper, we study the interplay between judgment and decision making in the context of data visualization. We first review how judgment and decision making have been conceptualized and studied in both visualization (Sec. \ref{sec:related_work}) and psychology research (Sec. \ref{sec:jdm_psychology}).
We next empirically study and discuss to what extent a judgment task (\emph{i.e.}, picking the optimal alternative based on expected values estimated visually) is a sufficient behavior predictor in a decision making task (\emph{i.e.}, to buy (or not) a player in fantasy sports). We examine judgments and decisions in-between conditions by casting participants in a series of either judgment or decision tasks with different visualizations (\condVisQDP, \condVisDensity, \condVisProbBar) and scenarios (\condScenSports, \condScenHumanitarian).
All our experimental material, data, analyses, and preregistrations can be found at the OSF link in the abstract.

\section{Related Work}
\label{sec:related_work}

\subsection{Judgment in Visualization}
\label{sec:related_work_judgment}

Visualization is used for decades to support various tasks that involve human \emph{judgments} for data exploration, analysis, and presentation. Judgment determines the need to build visualization systems: \emph{\quotes{If a fully automatic solution has been deemed to be acceptable, then there is no need for human judgment, and thus no need for you to design a vis tool.}\,\cite{Munzner2014Visualization}}.   
 
Visual judgments have been studied at the lower levels of processing, aiming to understand what people \emph{perceive} when looking at a visualization\,\cite{Elliott2020VisionScience}. 
 Examples include judging color discriminability\,\cite{Gramazio2016Colorgorical}, 
line orientation\,\cite{Talbot2012Empirical}, contrast\,\cite{Mittelstadt2014Contrast}, motion\,\cite{Veras2019Motion}, groupings \,\cite{Haroz2012Capacity}, 
 average positions\,\cite{Xiong2019Average}, extrema\,\cite{Patil2022}, and
 correlation patterns\,\cite{Harrison2014Correlation}.  
 
Higher-level judgments have been studied when looking at how users make sense of the displayed data -- that is, how they interpret the visualization to come to conclusions about the shown data\,\cite{Munzner2014Visualization,telea}. 
Examples include probability judgments  using intervals, density plots, quantile dotplots, and tables, which show sampled distributions of one\,\cite{Castro,Kale2021,Kayongo2022} or several attributes\,\cite{Zhang2015}; or more complex tasks such assessing causality\,\cite{Xiong2019Causality},
missing data\,\cite{Hayeong2019MissingData},
information credibility\,\cite{Schwarz2011}
or sensemaking\,\cite{Zhao2017Sensemaking}. 

Human judgments are further studied in the context of critical applications of visualizations, like medical diagnosis\,\cite{Borkin2011Diagnosis}
and weather forecasts\,\cite{Ferstl2016Forecast}.
Other judgment task examples include gauging the maintainability of a software system by visualizing its dependency graph\,\cite{holten06}; assessing traffic congestion of vessel fleets by visualizing their movements over space and time\,\cite{willems09}, and assessing the group structure of multidimensional samples to infer how easily classifiable a dataset is by machine learning\,\cite{rauber17}.
 
One can easily argue that most data visualizations involve a form of judgment -- indeed, if users were not able to, or interested in, making judgments about the depicted data values therein, that visualization would be useless.
 
\subsection{Decision Making in Visualization}
\label{sec:related_work_decisionmaking}
Following some authors, a key goal of visualization is to help \emph{\quotes{the decision maker to discover what should be said and done}}\,\cite{Bertin1983Semiology}, and that decision making is the most important reason of \emph{\quotes{why visualization is important}}\,\cite{Ward2015Interactive}. Visualization research invested much effort to study how users make decisions with visualized data\,\cite{Padilla2018,dimara2018task}. Decision tasks that have been studied include binary choices\,\cite{Kale2021}, such as to perform a humanitarian action or not\,\cite{Castro} 
or \emph{multi-attribute choice tasks} between many options\,\cite{Dimara2018DecisionSupport}.  
Other examples are choosing a holiday hotel based on price, hotel quality, landscape interest, and security level\,\cite{Dimara2018DecisionSupport} or choosing an investment based on company traits such as leadership ability, proprietary technology, market conditions, and competitor strength\,\cite{Zhang2015}. More examples of multi-attribute choice tasks cover time-interval choices (\emph{e.g.} when to arrive to catch a bus\,\cite{Fernandes2018QuantileDotplots}) and
group decision making\,\cite{Hindalong2022GroupDecisions}. 
Decision making has been further studied for critical applications such as finance\,\cite{Savikhin2008Applied},
hiring\,\cite{Narechania2021Lumos}, and 
engineering optimization decisions\,\cite{Cibulski2022Reflections}. 

Decision making is less studied than judgment in visualization. A recent review showed that decision tasks were involved in only 6\%  of quantitative and 4\% of qualitative evaluations of visualization tools for decision support, further attributed to the lack of relevant guidance from decision theory in visualization literature\,\cite{Dimara2021}.
Decision tasks are also omitted by visualization task taxonomies, including low-level and high-level ones\,\cite{Dimara2021}. Most real-world visualization applications involve domain experts who make judgments with data and almost never decision makers who make decisions based on data\,\cite{Dimara2021Organizations}.
Visualizations seemed to be used mostly to communicate a small part of the decisions already made by humans\,\cite{Dimara2021Organizations} or by AI systems\,\cite{cheng19,sperrle21}. Two recent reviews on the use of visualization for explainable AI (XAI)\,\cite{chatzimparmpas20,sperrle21} concluded that researchers study how to reveal the inner workings of a model (operations and outputs) and not how end-users ultimately involve the model in their decision making process. Meanwhile, the analysis of 940 tools reported by professional decision makers who described their work practices revealed the lack of a decision making tool that can support them through all steps of their decision making process,\cite{Dimara2021Organizations}. 

\subsection{Conflating Judgment and Decision Making}
\label{sec:related_work_conflation}

One barrier to studying decision making effectively is that judgments and decisions are terms not well-defined and separated. Visualization papers almost never formally define, or even casually describe, decision making \,\cite{Dimara2021}. Some authors consider decision making as a subpart of the high-level judgment task of sensemaking,\,\cite{Pirolli2005Sensemaking}; other authors see decision making as a broad expansive task that often contains a sensemaking subtask\,\cite{Dimara2021}; yet other authors view decision making and sensemaking as distinct tasks which should be supported by different visual analytics tools\,\cite{Sedig2012Interactivity}. Visual analytics literature often uses the term decision making to refer to algorithmic and not human decisions\,\cite{ieeevis}. 

This terminology confusion becomes more evident in empirical studies where the terms decision and judgment are used interchangeably.
For example, participant responses have been interpreted as \quotes{decisions}, though participants performed low-level visual judgment tasks, like spotting the higher bar between two alternatives\,\cite{Patil2022} and spotting the bar with the maximum average length between two bar sets\,\cite{Jardine2020}. When investigating if hypothetical outcome plots (HOPs) can facilitate people's judgment of trends, participants were asked to play the role of a newspaper editor whose job was to make a \quotes{decision} on the headline of a growth trend in the job market\,\cite{Kale2019}. The term decision was again used to refer to a visual judgment, although in that case, the confusion could have been influenced by the response type format -- the so-called  \quotes{two-alternative forced choice} (2AFC)-- and not by the nature of the task. 2AFC response formats though can be used for both judgment and decision tasks without changing the nature of the task. 
In another experiment, participants were asked to perform a decision task -- to issue (or not)  blankets to protect alpacas from cold\,\cite{Castro}. Yet, responses were this time interpreted as the ability of participants to make \quotes{judgments} on the probability of temperatures falling below 32 $^{\circ}$F\,\cite{Castro}. This conflation of terminology can also explain why the vast majority of visualization tools designed for decision support used solely judgment tasks in their evaluation\,\cite{Dimara2021}.

We are not aware of visualization studies that explicitly compare how people perform judgments and decisions. Two studies give some evidence on whether people are more accurate with judgments or decisions, as side notes within different experimental goals: Kale \emph{et al.}\,\cite{Kale2021} found that judgments and decisions can be better supported by different visualizations. Quantile dotplots supported better the judgment of the probability of superiority; interval and density plots supported better decision accuracy. Dimara \emph{et al.}\,\cite{Dimara2017a} studied how various narratives affect response accuracy, asking participants to select a house in a scatterplot based on price and size. Participants performed the task as a real estate agent, providing a judgment for their client; or as decision makers deciding which house to buy. 
Results showed that decision narratives elicited less accurate responses than judgment narratives.  
Interpreting this finding is hard. Both tasks were factually identical; an accurate response was a house that is not dominated by other houses that are both cheaper and bigger, and this accuracy criterion was explicitly communicated to both groups. In a realistic task, one possible interpretation could be that the decision group had subjective preferences of other salient features of the house, \emph{e.g.}, visual appearance. Yet, in this study, no other information was given besides size and price. 
Findings of the above two studies suggest that factors such as the visualization type or the task framing can elicit performance differences in judgment and decision making. 

To conclude, judgment and decision making are two terms conflated and two tasks which we do not yet know how they relate to each other.

\vspace{-1em}
\section{Judgment and Decision Making\\ in Psychology}
\label{sec:jdm_psychology}

There is often confusion about what to call judgment or decision (see Sec.\,\ref{sec:related_work}). To clarify this, we discuss \revision{definitions of judgment and decision making tasks in 
Sec.\,\ref{sec:jdmdefinitions}; we review empirical studies from psychology and economics in 
Sec.\,\ref{sec:jdmstudies}.}

\vspace{-1em}
\subsection{Judgment \& Decision Making Definitions}
\label{sec:jdmdefinitions}
\color{black}

The American Psychological Association (APA) dictionary of psychology defines judgment as \quotes{\emph{the capacity to recognize relationships, draw conclusions from evidence, and make critical evaluations of events and people; or the ability to determine the presence or relative magnitude of stimuli.}} Meanwhile, they define decision making as \quotes{\emph{the cognitive process of choosing between two or more alternatives, ranging from the relatively clear cut to the complex.}}\,\cite{APADictionary}. \revision{Extending the APA definitions, statistical decision theory\,\cite{berger2013statistical} posits that both judgments and subsequent decisions involve accounting for a certain level of uncertainty. However, unlike decisions, judgments do not require individuals to choose among alternatives\,\cite{eberhard2023effects}.}

Consistent with the APA definition, Diederich and Busemeyer\,\cite{diederich2013judgment} classify five judgment categories: (1) drawing conclusions from evidence (\quotes{Will you be in class on time?}; \quotes{What are the chances that you will get the attractive position you applied for?}); (2) making critical evaluations (\quotes{How much do you actually like that class?}); (3) value judgments (\quotes{A is interesting, beautiful, or better than or similar to B}); (4) category judgment (judging class membership), and (5) probability or quantity estimation. Meanwhile, Fischhoff and Broomell\,\cite{fischhoff2020judgment} define decision making as a process that consists of three parts: (1) judgment, \emph{i.e.}, the prediction of outcomes, (2) preference, \emph{i.e.}, how people weigh those outcomes, and (3) choice, \emph{i.e.}, how people combine judgments and preferences to make a decision. \revision{In line with Fischhoff and Broomell's definition\,\cite{fischhoff2020judgment}, in visualization research, Dimara and Stasko\,\cite{Dimara2021}, and Oral \emph{et al.}\,\cite{oral2023information}, argue that, unlike a judgment, a decision task should at the very least include a choice stage, following Simon's decision stages\,\cite{Simon1960ManagementDecision}.}

\revision{Despite the operational clarification on the need of a choice stage, the underlying mechanisms differentiating judgment from decision making remain elusive. Eberhard\,\cite{eberhard2023effects} concurs that decision making implies choosing actions with subsequent consequences, unlike judgment, which assesses alternatives without an obligation to act. Neuroimaging studies support this view by 
revealing activation in motor regions of the brain during decision tasks\,\cite{wispinski2020models}. This view is further reinforced by interviews with decision makers and analysts, as a participant summarizes: \quotes{A decision maker has to live with their decision while an analyst can just say what the best thing is and walk away!}\,\cite{Dimara2021Organizations}.
It is also shown that being an actor (\emph{vs.} an observer) enhances the sense of control and induces responsibility, attributes more pronounced in decision making\,\cite{Gold2015}. Beyond these factors, decisions pertain to the future\,\cite{karlsson1988phenomenological}, involve past experiences and personal identity\,\cite{karlsson1988phenomenological}, and stronger emotions\,\cite{pletti2017s} than judgments.} 

\revision{In summary, while decision making shares similarities with judgment, it embodies four distinguishing features: (\textbf{I}) it requires a choice among alternatives, implying a loss of the remaining alternatives, (\textbf{II}) it is future-oriented, (\textbf{III}) it is accompanied with overt or covert actions, and (\textbf{IV}) it carries a personal stake and responsibility for outcomes. The more of these features a judgment has, the more \quotes{decision-like} it becomes. When a judgment has all four features, it no longer remains a judgment and becomes a decision. This operationalization offers a fuzzy demarcation between judgment and decision making, in the sense that it does not draw a sharp line between the two concepts, but instead specifies the attributes essential to determine the extent to which a cognitive process is a judgment,  a decision, or somewhere in-between\,\cite{HAMPTON200679, decock2014graded}. We will use this operationalization in the rest of our article.}

\vspace{-1em}

\subsection{\revision{Judgment \& Decision Making Studies}}
\label{sec:jdmstudies}

\revision{When designing experiments}, psychology researchers appear more consistent with the aforementioned definitions when using the terms judgment and decision.
For instance, cultural psychology explicitly identifies judgment as the evaluation of the source of one's behavior (\emph{i.e.}, causal attribution), and it identifies as a decision whether to cooperate with or compete against a party (\emph{i.e.}, conflict decision)\,\cite{Savani2015}. However, we also observed that some psychology papers confuse judgment with decision making, just as in visualization research. For example, van Norman \emph{et al.}\,\cite{VanNorman2013} identified the visual judgment of a trend on whether an intervention improved students' performance or not as a decision.

\revision{As noted in Sec.\,\ref{sec:related_work_conflation}, 
with its inherent \emph{choice} framing, the use of AFC format sometimes leads researchers to categorize judgment tasks as decision tasks. 
However, in psychophysics\,\cite{gescheider2013psychophysics}, AFC is commonly used for low-level perceptual judgments like reporting the brightest stimuli\,\cite{liston2013saccadic}. Conversely, in behavioral economics, AFC is used for making decisions under uncertain conditions\,\cite{lynn2015decision}. Therefore, unlike the factors discussed in Sec.\,\ref{sec:jdmdefinitions}, the AFC format alone should not define the task as judgment or decision making.
} 

The psychology literature illuminated our discussion providing distinct definitions and more consistent use of the terms judgment and decision making. The domain of decision making under risk\,\cite{visschers2009probability} further provides insights into how to measure judgment and decision accuracy. This domain jointly studies judgments and decisions. Researchers studied judgment by measuring people's understanding of risks, \emph{e.g.}, perceived risk probabilities of different options\,\cite{chua2006risk}, conjointly with decision making by measuring risk behavior, \emph{e.g.}, willingness to pay for a product that reduces the risk of getting a disease or injury\,\cite{stone1997effects,chua2006risk}. Findings suggest that people tend to pay more for risk-reducing products when the risk is shown by icon arrays instead of numerical formats, claiming that icon arrays twist risk perception by highlighting the people at risk\,\cite{stone2003foreground}. In contrast, other studies showed icon arrays to improve judgments of health risk\,\cite{galesic2009using}. On the other hand, Wu \emph{et al.}\,\cite{Wu2017} showed that making accurate probability judgments did not improve search decisions. Likewise, although people can judge doing something as morally wrong, they may decide to behave in a way that is not consistent with their judgment\,\cite{Gold2015}.

Although judgment and decision accuracy have not been explicitly contrasted in any of these studies, we saw that is plausible that judgments can influence decisions in all possible directions. We saw accurate judgments which improve or do not improve decisions, as well as inaccurate judgments that even lead to better decisions. It thus remains unclear if judgment accuracy can guarantee better subsequent decisions. One possible confounding factor in all previous studies is that judgments and decisions were tested in within setups. So, we cannot exclude the possibility that asking a person to perform a judgment before a decision can itself influence the decision behavior.

To conclude,  we are still not aware of any attempt that explicitly studies the difference in response accuracy between judgment and decision making.

\vspace{-0em}

\section{Research Hypotheses}
 
 \label{sec:researchHypothesis}

To empirically investigate the interplay between judgment 
and decision making, we formulated the following hypothesis
which \revision{was preregistered prior to data collection at the OSF link given at the end of the abstract.}

\begin{itemize}
    \setlength\itemindent{-0.6em}
    \item[\textbf{Hr}]: Performance on a judgment task with a data visualization is not a good proxy for performance with the corresponding decision task with the same data visualization.
\end{itemize}

\noindent Hypothesis Hr was assessed through 5 sub-hypotheses:

\begin{itemize}
    \setlength\itemindent{-0.2em}
    \item[\textbf{Hr${_1}$}]: Decision accuracy overall differs from judgment accuracy. This hypothesis does not assume directionality because there are conflicting trends in related works (discussed in Secs.\,\ref{sec:related_work_conflation} and \ref{sec:jdm_psychology}).
    
    \item[\textbf{Hr${_2}$}]: Decision accuracy and judgment accuracy differ in their sensitivity to the underlying scenario. This hypothesis is motivated by work suggesting that the prospective death of an animal elicited irrational responses in decision accuracy (\emph{e.g.}, always issuing blankets to the Alpacas regardless of the weather forecast probabilities)\,\cite{Castro}.
    
    \item[\textbf{Hr${_3}$}]: There is a specific visualization where decisions elicit different accuracy than judgments.
    
    \item[\textbf{Hr${_4}$}]: There is a specific visualization and scenario where decisions elicit different accuracy than judgments.
    
    \item[\textbf{Hr${_5}$}] :The \DT differs from the \JT in its vulnerability to the sub-optimal heuristic strategies \quotes{risk-aversion} and \quotes{status-quo} biases. This hypothesis is motivated by research showing decisions are hindered by loss aversion and avoidance of changing the person's current state (status-quo)\,\cite{dimara2018task}.
    
\end{itemize}

\noindent Our hypotheses were hierarchically structured. The \JT is not expected to be a sufficient proxy for the \DT [Hr] because either of the following: (a) decisions elicit overall different accuracy from judgments [Hr${_1}$]; (b) \DA overall differs from \JA in its sensitivity to the underlying scenarios [Hr${_2}$]; (c) \DA is better facilitated by different visualizations [Hr${_3}$]; (d) there is at least a specific visualization and scenario where decisions elicit different accuracy from judgments [Hr${_4}$]; (e) \DT differs from \JT in its vulnerability to suboptimal heuristic strategies, \emph{e.g.}, risk-aversion and status-quo biases [Hr${_5}$]. Support for at least one of the sub-hypotheses [Hr${_1}$ - Hr${_5}$] provides support for the main hypothesis [Hr]. However, the larger the number of the supported sub-hypotheses are, the stronger the understanding of the nature of unsuitability of a \JT as a proxy for \DT will be.  
Before we provide the statistical hypotheses (Sec.\,\ref{sec:exp1-statisticalHypotheses}) which map to the above research hypotheses, we detail our experiment design.  
\vspace{-1em}

\section{\revision{Design Rationale: Judgment Vs. Decision }}\label{sec:design_rationale}

\revision{To investigate \textbf{Hr}, we assigned participants to two tasks: judgment and decision. We next outline our design choices for these tasks based on insights from Secs.\,\ref{sec:related_work_conflation}, \,\ref{sec:jdmdefinitions}, and \ref{sec:jdmstudies}.}

\subsection{\revision{Differences Between Tasks}}
\label{sec:design-rationale-differences}

\revision{
\noindent \textbf{Question Type:} As illustrated in Fig.\,\ref{fig:tasks}, the sole distinction between the two tasks lay in their respective \emph{framings}: For the judgment task, we asked the question \quotes{What is the best option?} (an observation framing); for the decision task, we asked \quotes{What do you choose?} (an action-oriented framing). This framing manipulation is almost identical to the narrative study by Dimara et al.\,\cite{Dimara2017a} discussed in 
Sec.\,\ref{sec:related_work_conflation}, which designated participants as either real-estate analysts who provide judgments or house buyers who make decisions. Drawing from the insights in Sec.\,\ref{sec:jdmdefinitions}, we hypothesized that the action-oriented framing (decision feature \textbf{III}) of the decision task would elicit a heightened sense of control, responsibility (decision feature \textbf{IV}), and consequently emotion, while the judgment framing would position participants more as analytical observers.}

\subsection{\revision{Identical Design Choices Between Tasks}}
\label{sec:design-rationale-common}

\revision{
In order to ensure validity in comparing the judgment and decision tasks and minimize confounding variables, we maintained uniformity in the remaining parameters:}%

\vspace{0.4em}
\noindent \revision{\textbf{Response Type:}
Both tasks employed the AFC format. As mentioned in Sec.\,\ref{sec:jdmstudies}, this format is apt for both judgment-related psychophysics and economic decision making. Unlike the binary choices (2AFC) used in studies\,\cite{Castro,Kale2021}, we adopted 5AFC to increase the resolution of the accuracy score, which would otherwise be limited to 0 or 1 (a 50\% chance to identify the best outcome by pure chance). Both tasks featured identical option sets: one optimal, one risk-averse, one status-quo (\textbf{Hr${_5}$}), and two alternative options, detailed in Sec.\,\ref{sec:exp1-dataset}.}

\vspace{0.4em}
\noindent \revision{\textbf{Response Accuracy:}
We measured the accuracy of both tasks using the same expected value metric, as detailed in Sec.\,\ref{sec:exp1-measures}.}

\vspace{0.4em}
\noindent \revision{\textbf{Task Context:}
All participants viewed three visualizations (\textbf{Hr${_3}$}) across two scenarios (\textbf{Hr${_2}$}): one on sports and another, emotionally charged, on children in war. We designed these scenarios to maintain a consistent emotional influence in both tasks, regardless of question framing. See Sec.\,\ref{sec:exp1-scenarios} for scenario details and Sec.\,\ref{sec:exp1-visualizations} for the visualizations.}

\vspace{0.4em}
\noindent \revision{\textbf{Incentives and Instructions:}
We compensated all participants and encouraged \quotes{as accurate as possible} responses. While many experimental designs, particularly in economics, introduce extrinsic motivation through performance-based incentives to emulate real-life decisions, we designed our scenarios to foster intrinsic motivation, consistent with psychology study methodologies. We opted against performance-based incentives due to concerns raised in\,\cite{Kale2021} preventing disparities between the judgment and decision tasks that might skew our comparative analysis. Guiding participants explicitly toward a \quotes{correct} way of answering would counter our objective of capturing innate strategies in judgments and decisions. In real-world situations, the distinct incentive structures for genuine decisions and judgments would make direct comparisons even more challenging. Thus, although we trained participants on probabilities, visual interpretation, and cost-profit trade-offs, we refrained from priming them with specific \quotes{correctness} benchmarks or linking them to incentives other than the ones implied in the task framing.}

\vspace{-1em}

\section{Experiment 1}
\label{sec:exp1}

This experiment tested our hypothesis \textbf{Hr} via sub-hypotheses \textbf{Hr${_1}$ - Hr${_5}$}. We exposed participants to visualizations: \condVisQDP, \condVisDensity, and \condVisProbBar, asking questions involving probability and cost/profit estimations. We examined the influence of the task ( \JT  \emph{vs.} \DT), scenario (\condScenHumanitarian \emph{vs.} \condScenSports), and visualization on response accuracy. All materials and preregistration details are at the OSF link in the abstract.

\begin{figure}
\centering
\includegraphics[width=1\linewidth]{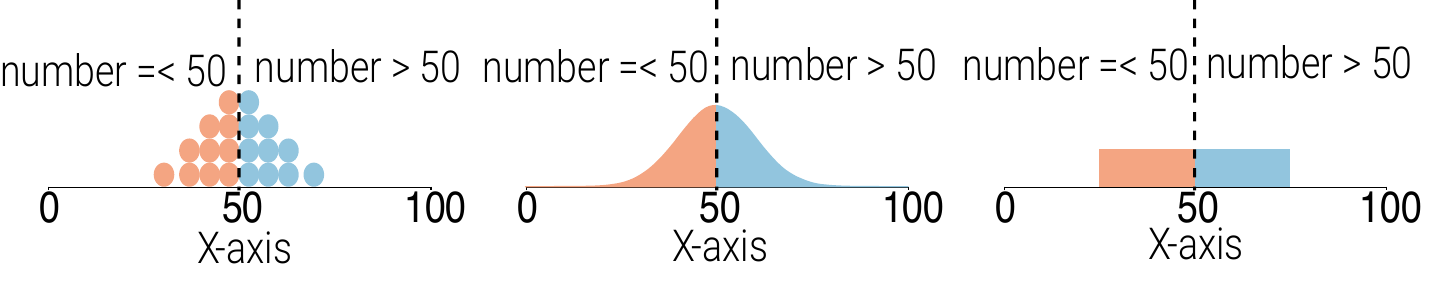}
\vspace{-2em}
\caption{The three visualizations used in both experiments:  quantile dotplot (left), density plot (middle), and probability bar (right).}
\vspace{1em}
\label{fig:visualizations}
\end{figure}

\subsection{Visualizations}
\label{sec:exp1-visualizations}

We utilized three uncertainty visualizations: \condVisQDP, \condVisDensity, and \condVisProbBar, as shown in Fig.\,\ref{fig:visualizations}. Our choices were based on studies that explored decision making and probability estimates with uncertainty visualizations\,\cite{Castro, Kale2021}. Unlike\,\cite{Castro, Kale2021}, which used intervals, we opted for \condVisProbBar visualization. While the dot count in \condVisQDP and area in \condVisDensity convey probability information, interval bar widths don't directly represent probabilities but denote 95\% of possible outcomes. We therefore adapted the interval's horizontal bars to incorporate exact probabilities, ensuring they align in width with \condVisQDP and \condVisDensity.

\vspace{-1 em}
\begin{figure}
\vspace{-2 em}
\centering
\includegraphics[width=1\linewidth]{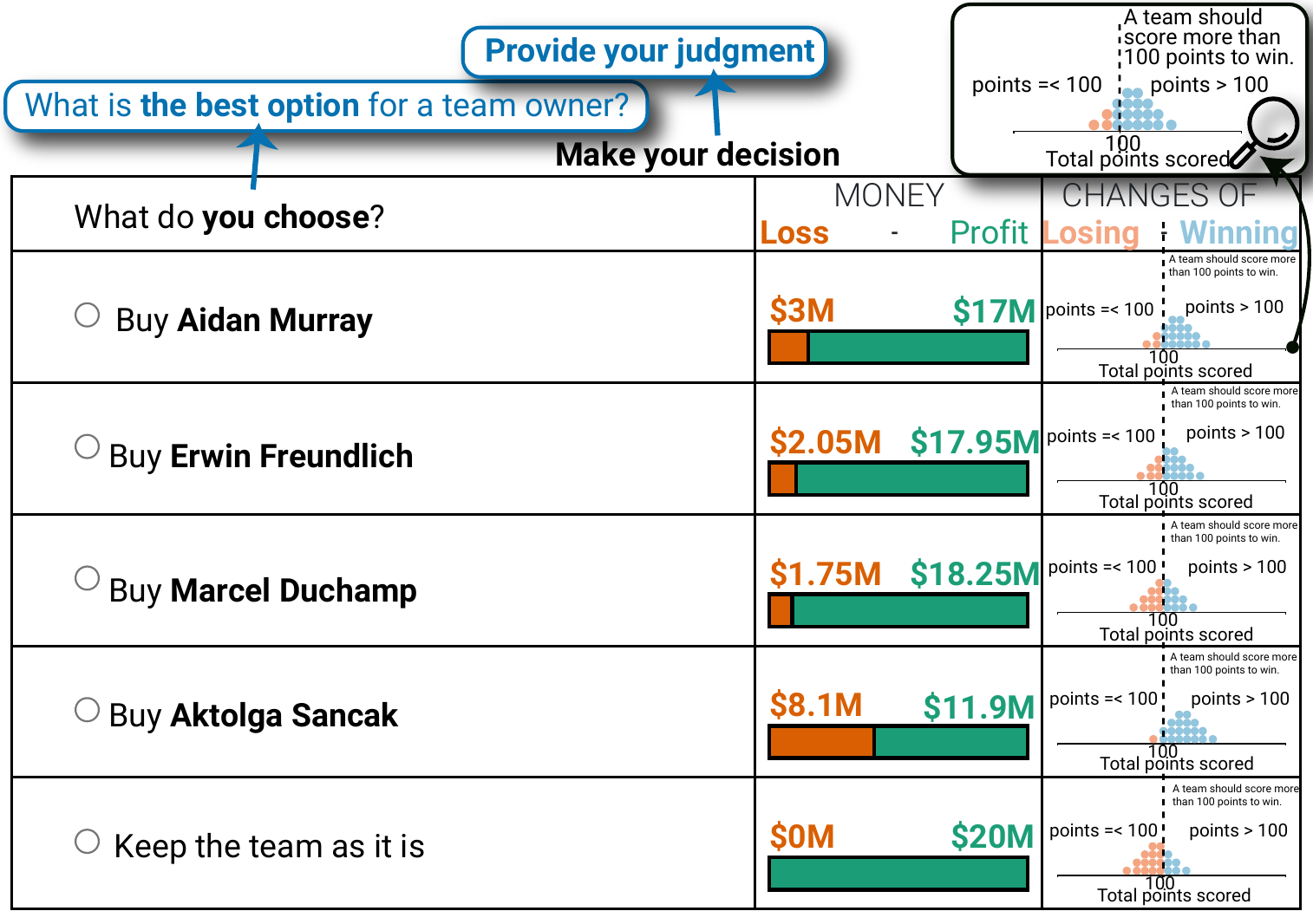}
\vspace{-2em}
\caption{\revision{Example of the \DT condition with the sports scenario and quantile dotplot visualization. The \JT version was identical, except for modifications in the title and question, highlighted in blue.}}
\vspace{-1.5em}
\label{fig:tasks}
\end{figure}

\subsection{Scenarios and Stimuli}
\label{sec:exp1-scenarios}
We showed two scenarios, \condScenSports and \condScenHumanitarian (\textbf{Hr${_2}$}). 
In the \condScenSports scenario, adapted from Kale \emph{et al.},\cite{Kale2021}, a team owner can buy players to create a team.
Each player has a cost (shown with red bars in Fig.\,\ref{fig:tasks}),
but if they help their team to win, the team gets a profit (shown with green bars in Fig.\,\ref{fig:tasks}).
The winning probability of a player, displayed in blue in Fig.\,\ref{fig:visualizations}, is contingent upon a certain threshold of points the team needs to score for victory (\emph{e.g.}, 100).
The decision task was to choose which player to buy (Fig.\,\ref{fig:tasks}). 
The judgment task was to indicate which player is the best for the team owner (Fig.\,\ref{fig:tasks} \revision{blue highlights}). 
Both tasks had 5 alternative responses, 4 different players, or the option to do nothing (for Hr${_5}$). 
The best response was the one that maximizes the expected value (see Sec.\,\ref{sec:exp1-measures}). Colors were chosen with ColorBrewer to ensure safety for color-blind viewers.

For the second scenario, we sought a more affective context than \condScenSports, such as aiding Alpacas in Castro \emph{et al.}\,\cite{Castro} and humanitarian visualizations\,\cite{Morais2021Anthropographics}. However, it was essential that the storyline supports the same components as the \condScenSports scenario, including 5 alternatives, cost-profit considerations, probability trade-offs, and a do-nothing option. 
We adopted a narrative involving a war between two fictitious countries, Syldavia and Borduria (names from the Adventures of Tintin). 
Orphans in Syldavia need essential goods. Syldavia can either wait for aid from neighboring countries without cost (do-nothing option) or fund one of four humanitarian organizations to help the orphans. Each organization has a transportation fee, charged from Syldavia's limited resources. If they succeed in delivering, the kids receive the goods (profit). Each organization has its own delivery probability before borders close, as each player has varied the probability of the win in Fig.\,\ref{fig:tasks}. If unsuccessful, Syldavia's money is wasted. 
Again, the decision was to choose one of the 4 organizations or do nothing, and the judgment was to indicate the best option for Syldavia. 
Except for text, the stimulus of \condScenHumanitarian was identical to the \condScenSports scenario. Detailed experiment instructions are in the supplementary materials.

\vspace{-1.0em}
\subsection{Dataset}
\label{sec:exp1-dataset}
We generated synthetic data for probability and cost/profit information. We selected 33\% and 90\% as the probability endpoints; 33\% and 90\% were always for status-quo and risk-averse alternatives, respectively. Since the brain represents probability on the log-odds scale\,\cite{zhang2012ubiquitous}, we converted the endpoints into log-odds units. Then, we sampled on this logit-transformed scale using linear interpolation between the endpoints.
We created three difficulty levels, \emph{i.e.}, easy, medium, and hard. 
At the \emph{easy} level, in addition to status quo and risk-averse alternatives, we added the best alternative and two alternatives dominated by the best alternative. While one dominated alternative had the same cost and profit value as the best alternative, its probability of winning was lower. 
The other dominated alternative had the same probability but more cost and less profit compared to the best. 
At the \emph{medium} level, in addition to status quo and risk-averse alternatives, we added the best alternative, a second alternative, and an alternative dominated by both the best and the second alternatives. 
At the \emph{hard} level, in addition to the status quo and risk-averse alternatives, we added the best alternative and two others. At the easy and medium levels, accuracy scores are 0, 0.25, 0.50, 0.75, and 1 for the risk-averse, status-quo, alternative 1, alternative 2, and best alternative, respectively. 
On the other hand, at the hard level, the accuracy of different alternatives changed. The best alternative and alternative 2 still have accuracy scores of 1 and 0.75. But, the status-quo, alternative 1, and risk-averse alternatives have accuracy scores of 0, 0.25, and 0.50, respectively.
All cost and profit values were randomly selected and fixed considering the defined probability and accuracy values. The dataset is available at the OSF link.

\subsection{Procedure}
\label{sec:exp1-procedure}

Participants completed a consent form and underwent training on the visualizations involved, \condVisDensity, \condVisProbBar, and \condVisQDP, and cost/profit bars. During the training, they tackled two types of tasks for each visualization. The first task involved identifying the alternative with the highest probability depicted by the visualization among five alternatives. The second task was selecting the probability range depicted by the visualization. \minorrevision{Finally, we instructed participants on how to read cost and profit bars by explicitly stating that they should consider the length of the green bar in conjunction with the probability of success and the length of the orange bar in conjunction with the probability of failure (e.g., see intro1-page74 under experiment\_screens on OSF Supplementary materials)}

Post-training, participants performed either judgment or decision tasks, based on their assigned condition, with all scenarios and visualization types. They then completed a demographics questionnaire, including gender, age, education level, and country of residence, an Adaptive Berlin Numeracy Test (BNT),\cite{Cokely2012}, and answered two optional questions about their strategy and comments. On average, the experiment lasted approximately 30 minutes.
Details on the procedure are in the supplementary materials.

\vspace{-1em}

\subsection{Experiment Design}
\label{sec:exp1-experimentDesign}

The experiment utilized a mixed design. Scenario (\condScenSports and \condScenHumanitarian) and visualization (\condVisDensity, \condVisProbBar, and \condVisQDP) were within-subjects independent variables, ensuring participants encountered all scenarios and visualizations.
The task (\JT or \DT) was a between-subjects independent variable; thus, a participant, randomly assigned,  either provided judgments or made decisions.
The motivations for choosing a between-subjects design is detailed in Sec.\,\ref{sec:jdmstudies}.
 The order for within-subjects conditions was entirely randomized, while difficulty sequence remained fixed (easy, medium, hard as discussed in Sec.\,\ref{sec:exp1-dataset}).
Overall, each participant tackled 18 trials (3 visualizations x 2 scenarios x 3 difficulty levels).

\vspace{-1em}

\subsection{Measures}
\label{sec:exp1-measures}
We used the following measures: 

\smallskip
\noindent\emph{Accuracy} as \revision{the proportion of the difference between the expected value (EV) of a participant's choice (among five different alternatives) and the choice with the worst possible EV, relative to the difference between the choices having the best and worst possible EVs:}  
\[
    Accuracy = \frac{|EV_{choice} - EV_{worst}|}{|EV_{best} - EV_{worst}|}.
\]

\noindent \minorrevision{To illustrate this with an example: assume EVs are placed on a numerical axis. The accuracy score is calculated based on how far the EV of the chosen option is from the worst EV, scaled by the range of the numerical axis (i.e., the difference between the best and worst values). In this case, if a participant chooses the best option, the accuracy score is 1 while it is 0 if the worst option is chosen. The EV value of the options in the middle (i.e., the other three options) is adjusted in such a way that each option is equidistant from the adjacent ones. This results in a set of possible accuracy scores: \{0, 0.25, 0.5, 0.75, 1\}, as explained in Sec.\,\ref{sec:exp1-dataset}.}

\smallskip
\noindent\emph{Time} of completion in seconds. 

\smallskip
\noindent\emph{Risk Literacy} with the Adaptive Berlin Numeracy Test\,\cite{Cokely2012}.

\vspace{-1em}
\subsection{Statistical Hypotheses}
\label{sec:exp1-statisticalHypotheses}

We next translate our research hypotheses (Sec.\,\ref{sec:researchHypothesis}) to the following statistical (thus, testable) hypotheses:
\begin{itemize}
    \setlength\itemindent{-0.6em}
    \item[H${_1}$]: The mean accuracy score of the decision task across scenarios and visualizations is measurably different than the mean accuracy score of the judgment task.
    \item[H${_2}$]: The mean accuracy score of the decision task is measurably different than the mean accuracy score of the judgment task with scenario $S$, where $S \in$ \{\condScenSports, \condScenHumanitarian\hspace{-.35em}\}.
    \item[H${_3}$]: The mean accuracy score of the decision task is measurably different than the mean accuracy score of the judgment task with visualization $V$, where $V \in$ \{\condVisQDP, \condVisDensity, \condVisProbBar\hspace{-.35em}\}.
    \item[H${_4}$]: The mean accuracy score of the decision task is measurably different than the mean accuracy score of the judgment task in scenario S and visualization V, where S $\in$  \{\condScenSports, \condScenHumanitarian\hspace{-.35em}\}, and V $\in$ \{\condVisQDP, \condVisDensity, \condVisProbBar\hspace{-.35em}\}.
    \item[H${_5}$]: In the \DT, the \quotes{status quo} alternative is chosen measurably more or less frequently than it is done in the \JT -- and the same for the \quotes{risk-averse} alternative in the \DT \emph{vs.} the \JT. Both expectations are on \quotes{more.}
\end{itemize}

\vspace{-1em}
\begin{figure}[t]
\centering
\includegraphics[width=1\linewidth]{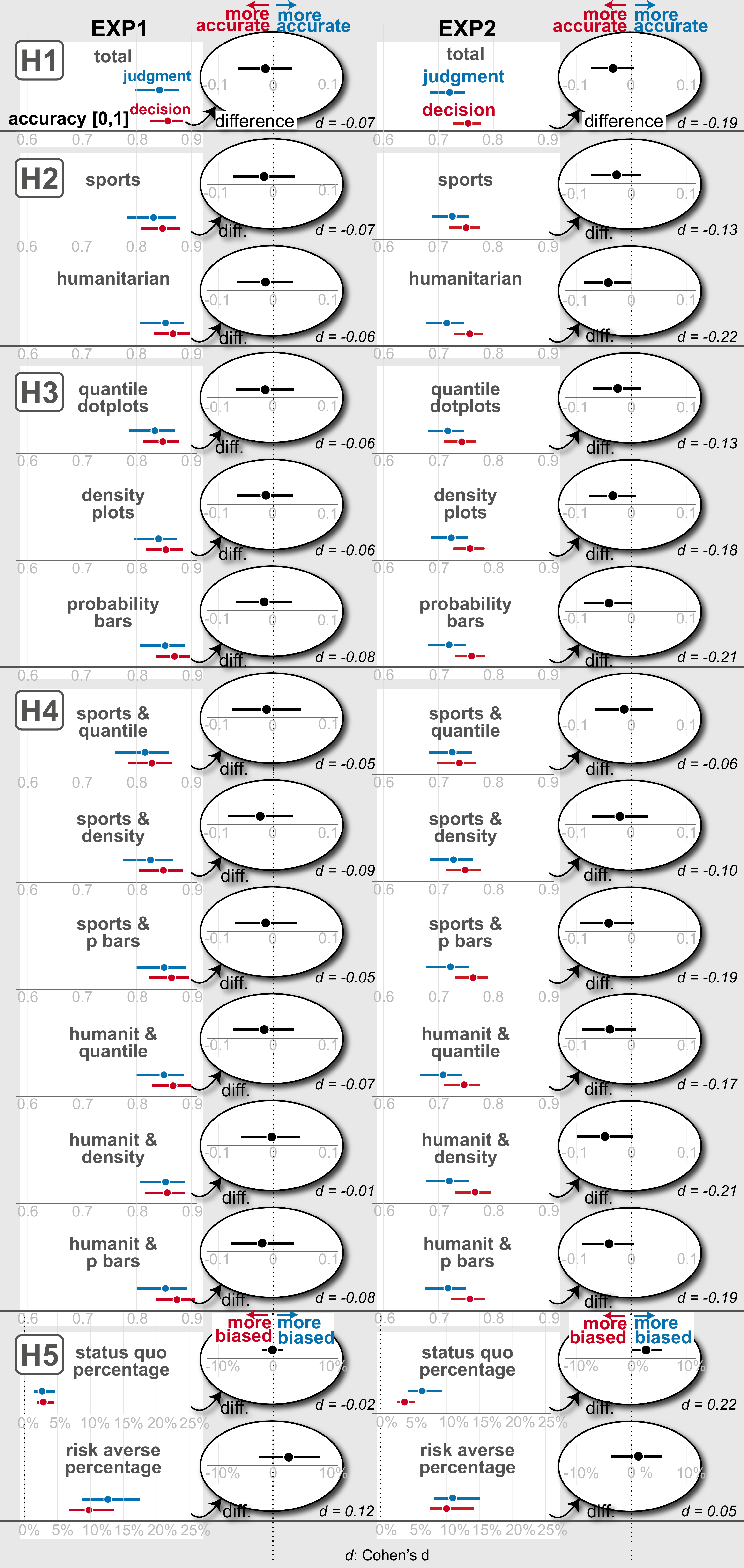}
\vspace{-2em}
\caption{Results for H${_1}$ - H$_5$ showing CI of average accuracy and bias percentage per condition for experiments 1 (left) and 2 (right). 
Ellipses depict the difference (black CI)  between average judgment (blue CI) and decision (red CI)  accuracies (for H${_1}$ - H${_4}$) and bias percentage (H${_5}$). \revision{Cohen's d values for each difference are bottom-right of the ellipses.} Chart titles (\emph{e.g.}, total, sports) indicate the specific condition.}
\vspace{-2em}
\label{fig:mean_accuracy}
\end{figure}

\begin{figure}
\centering
\includegraphics[width=1\linewidth]{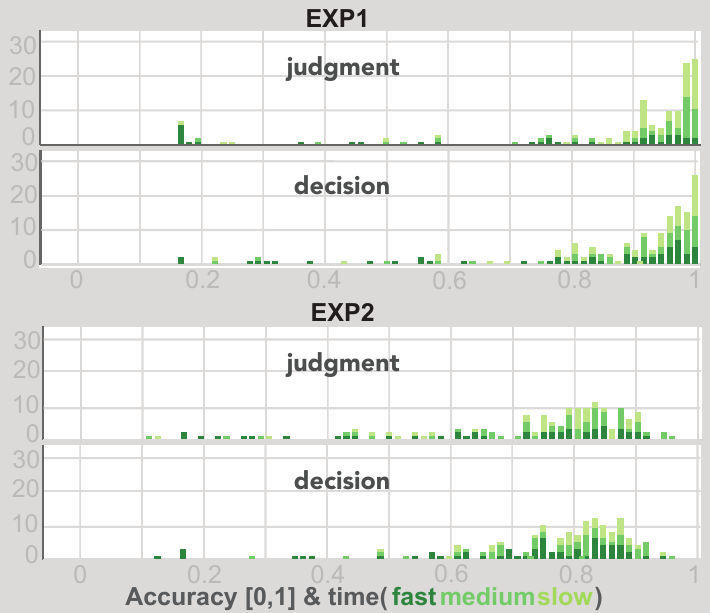}
\vspace{-2em}
\caption{
Accuracy scores ($x$ axis) \emph{vs.} participant count ($y$ axis) colored by response time for both experiments. Log-transformed time ranges:\revision{ fast (0.52-1.43], medium (1.43-1.74], slow (1.74-3.04], corresponding to minutes: (1.68-4.18], (4.18-5.70], (5.70-20.90]}.
}
\vspace{-1.5em}
\label{fig:raw_accuracy}
\end{figure}

\subsection{Participants}
\label{sec:exp1-participants}

We secured ethics, privacy, and data management approval for our study from our department's respective committees. Participants were sourced from the Prolific crowd-sourcing platform (platform specifics: all countries, standard sample, prior minimum approval rate: 95\%, payment: 6 pounds/hour). Our target sample size was $N = 300$ (around 150 per task) following previous studies,\,\cite{Castro, Kale2021}.
Of 162 participants in the \JT condition who consented, 16 quit before completing, and 4 were excessively slow (see preregistered exclusion criteria). From 171 participants in the \DT condition who consented, 13 didn't finish. Thus, our final count was 300 participants ($N = 142$ for \JT and $N = 158$ for \DT), with 62\% male and 51\% aged 18-30, representing 24 countries, primarily Europe (main countries: UK 33\%, Poland 14\%, South Africa 11\%).

\vspace{-1em}
\subsection{Results of Experiment 1}
\label{sec:exp1-resultsOfExperiment1}
We planned and preregistered all analyses (see OSF link in abstract) prior to data collection. We used an \emph{estimation approach} to statistical reporting, \emph{i.e.}, we base all our inferences on confidence intervals reported graphically, seeing statistical evidence as lying on a continuum rather than being binary\,\cite{cumming2014new,dragicevic2016fair,besanccon2019continued}. For guidance on reading graphs with confidence intervals and relate them to p-values, \revision{see\,\cite{cumming2005inference,krzywinski2013error}.}

All our confidence intervals (CIs) are 95\% BCa bootstrap confidence intervals\,\cite{kirby2013bootes}. Our CIs are not corrected for multiplicity. Thus, since we test five hypotheses and some of them break down into multiple statistical hypotheses (up to 6), any isolated finding must be taken as \revision{tentative}. As a reminder, consistent with our statistical hypotheses H${_1}$ -- H${_4}$ (Sec.\,\ref{sec:exp1-statisticalHypotheses}), we expect to find accuracy differences between the \JT and the \DT across scenarios and visualizations. Also, consistent with H${_5}$, we expect the percentage of status-quo and risk aversion biases to differ across the \JT and the \DT. 

The main results of experiment 1 are shown in Fig.\,\ref{fig:mean_accuracy} (left column). The point estimates and CIs for differences (in the ellipses on Fig.\,\ref{fig:mean_accuracy}) were neither planned nor preregistered and were computed later to facilitate the interpretation of statistical evidence. To further assist the interpretation of effect magnitudes, we also chose to report standardized effect sizes (Cohen's $d$), of which computation was also neither planned nor preregistered. 

For  H${_1}$, we compared the mean accuracy score between the \JT and the \DT. Results are reported graphically in Fig.\,\ref{fig:mean_accuracy}, column EXP1, row H1. \revision{We found no evidence for a difference between \JA and \DA.} Likewise H${_1}$, the results were also inconclusive for hypotheses from H${_2}$ to H${_5}$ (see corresponding rows in Fig.\,\ref{fig:mean_accuracy} for visual comparison). 

\vspace{-1em}

\subsection{Additional Analyses}
\label{sec:exp1-additionalAnalyses}
\subsubsection{Response time and task accuracy}
\label{sec:exp1-responseTimeAndTaskAccuracy}
We measured the possible correlation between response time and task accuracy to see whether the speed of responses can relate to what people decide or judge.
Correlation analysis showed that there is some evidence for a positive trend between response time and both \JA, \emph{r} = 0.30, CI [0.12, 0.44], and \DA, \emph{r} = 0.18, CI [0.03, 0.31]. Yet, point estimates for the \emph{r} values were small.

\subsubsection{Risk literacy and task accuracy}
\label{sec:exp1-riskLiteracyAndTaskAccuracy}
We examined how risk literacy, intended to capture people's ability to accurately calculate and understand probabilities,  relates to \JA and \DA.  
We found no clear evidence between risk literacy and \JA (\emph{r} = 0.09, CI [-0.08, 0.24]) and between risk literacy and \DA (\emph{r} = 0.19, CI [0.04, 0.33]). Yet, if there is any, it is more likely that more accurate decisions can be associated with higher risk literacy.

\vspace{-1em}
\subsection{Summary of Experiment 1}
\label{sec:exp1-conclusionsOfExperiment1}

The results were inconclusive for all our hypotheses on whether there is a performance difference (or not) between conditions.  We found no evidence of \JA-\DA differences across scenarios, visualizations, or any specific scenario-visualization pair. Also, we found no evidence for status-quo and risk-aversion bias percentage differences between conditions.

\vspace{-1em}

\section{Experiment 2}
\label{sec:exp2}

The lack of observed \JA $-$ \DA  difference in experiment 1 could be, we believe, a ceiling effect, as many accuracy scores were equal to 1 (see raw accuracy scores, Fig.\,\ref{fig:raw_accuracy} top). 
Hence, we conducted a second experiment increasing task difficulties [\textbf{C1}]. Given participants' comments as \quotes{\emph{Yes, the study wasn't randomized so after a while it was easy to remember which options I chose previously even though they were shuffled around or the scenario was changed}} and \quotes{\emph{The study was enjoyable, but I would say I found the sections checking understanding at the start were a bit longer than I think they needed to be}}, 
we increased the options' variety across trials [\textbf{C2}] and  
shortened the tutorial length [\textbf{C3}].  

\vspace{-0.5em}
\subsection{Scenarios, Tasks and Stimuli}
\label{sec:exp2-scenarios}

Experiment 2 was identical to experiment 1, including hypotheses and design, except for the changes described next. Scenarios, tasks, and stimuli were identical except for a few small but important changes in experiment instructions: To ensure that the judgment and decision framing did not escape participants' attention, we emphasized it further by changing \quotes{What is the best option for Syldavia?} to \quotes{Consider that this is your judgment. What is the best option for Syldavia?},
and \quotes{What is the best option for a team owner?} to \quotes{Consider that this is your judgment. What is the best option for a team owner?} for the judgment task. 
We also changed \quotes{What do you choose?} to \quotes{Consider that this is your decision. What do you choose?} for the decision task. Also, we changed \quotes{Next you will be asked to provide judgments.} to \quotes{Next you will be asked to provide your own judgments.} and \quotes{Next you will be asked to make decisions.} to \quotes{Next you will be asked to make your own decisions.} To avoid shortening the experiment duration if participants felt they needed to make computational calculations on the side, we also clarified that \quotes{This experiment is not a math test. No calculations are required on your side, but your answer should reflect your judgment/decision as fast and accurately as possible.} 

\subsection{Dataset}
\label{sec:exp2-dataset}
The dataset generation algorithm was identical to experiment 1 (Sec.\,\ref{sec:exp1-dataset}), including  difficulty levels (easy, medium, hard). However, there was no option with the same probability or cost/profit value, making the comparison of options harder than experiment 1 [\textbf{C1}]. Also, the best alternative did not always have the second-best probability; for example, it had the third and fourth-best probabilities at medium and hard levels, respectively. In addition, another difference was that we created unique probability and cost/profit pairs for each trial and that we selected three sets, \emph{i.e.}, Set1:[25\%, 90\%], Set2: [30\%, 90\%], Set3: [33\%, 90\%] for the probability endpoints to increase the variety of trials  [\textbf{C2}]. Like in experiment 1, we transformed the endpoints of each set into log odds and applied linear interpolation to sample three other probability values between the endpoints of each set.
The dataset is available at the OSF link at the end of the abstract.

\vspace{-0.5em}
\subsection{Procedure}
\label{sec:exp2-procedure}
The procedure was identical to experiment 1 (Sec.\,\ref{sec:exp1-procedure}), 
except for decreasing the length of training [\textbf{C3}]. Participants  answered only the question of finding the highest probability among 5 alternatives until they found the correct answer or up to 5 trials in total, and did not fill a literacy test.

\subsection{Measures}
\label{sec:exp2-measures}
The measures were identical to experiment 1 (Sec.\,\ref{sec:exp1-measures}), except removing the risk literacy measure to reduce duration given the lack of observed relation between risk literacy and accuracy  (Sec.\,\ref{sec:exp1-riskLiteracyAndTaskAccuracy}).

\subsection{Participants}
\label{sec:exp2-participants}

We again recruited participants via Prolific.  Sample specifications, payment, planned sample size, and exclusion criteria were as in experiment 1 (Sec.\,\ref{sec:exp1-participants}). We got valid data from 301 participants ($N = 146$ for the \JT; $N = 155$ for the \DT), 61\% male, 56\% aged between 18-30, from 26 countries (top ones: UK 32\%, USA 15\%, Poland 10\%, and South Africa 10\%).

\subsection{Results of experiment 2}
\label{sec:experiment2_results}

All analyses were planned and preregistered (see OSF link in abstract) prior to data collection. 
Fig.\,\ref{fig:mean_accuracy} (right) shows the results. Analysis methods, computations, and their explanation are identical to experiment 1.

\subsubsection{Statistical hypothesis H${_1}$} 
We compared the mean accuracy score of the \JT and the \DT (see Fig.\,\ref{fig:mean_accuracy}, column EXP2, row H1). We found \minorrevision{suggestive evidence for a difference between \JA and \DA}, more specifically, that \DA is higher than \JA.

\subsubsection{Statistical hypothesis H${_2}$}
We compared the mean accuracy score of the \JT and the \DT across scenarios. In the \condScenSports scenario (see Fig.\,\ref{fig:mean_accuracy}, column EXP2, row H2,  sports), we \revision{did not find evidence that \JA and \DA differ}. However, for the \condScenHumanitarian scenario 
(see Fig.\,\ref{fig:mean_accuracy}, column EXP2, row H2, humanitarian), we found \minorrevision{suggestive evidence for a difference between \JA and \DA}, more specifically, that \DA is higher than \JA in the \condScenHumanitarian scenario.

\subsubsection{Statistical hypothesis H${_3}$}
We compared the mean accuracy score of the \JT and the \DT across visualizations, \emph{i.e.} quantile dotplots, density plots, and probability bars. We found \revision{suggestive evidence for a difference between \JA and \DA in probability bars}, but \revision{the evidence is weaker for the remaining visualizations} (see Fig.\,\ref{fig:mean_accuracy}  
column EXP2, row H3). 

\subsubsection{Statistical hypothesis H${_4}$}
We compared the mean accuracy of the \JT and the \DT across scenarios and visualization pairs. Results were inconclusive for the \condScenSports scenario, except when the probability bars were used 
(see Fig.\,\ref{fig:mean_accuracy}, column EXP2, row H4). For the \condScenHumanitarian scenario, we \revision{found suggestive evidence for a difference across all three visualizations} 
(see Fig.\,\ref{fig:mean_accuracy}, column EXP2, row H4). 

\subsubsection{Statistical hypothesis H${_5}$}
We compared the mean status quo and risk-aversion percentage for the \JT and the \DT. We \revision{did not find evidence for a difference between the \JT and the \DT on risk-aversion percentage} 
(see Fig.\,\ref{fig:mean_accuracy}, column EXP2, row H5, risk averse percentage). In contrast, we \revision{found suggestive evidence for a difference between the \JT and the \DT} on status quo percentage. Specifically, there was more status-quo bias in the \JT than in the \DT 
(see Fig.\,\ref{fig:mean_accuracy}, column EXP2, row H5, status quo percentage).

\vspace{-1em}

\subsection{Additional Analyses}

We investigated if there is a correlation between \JA and \DA and response time and found some evidence for a moderate positive correlation, \emph{r} = 0.29, CI [0.12, 0.43]. We also found some evidence that increasing response time may lead to more accurate judgments, \emph{r} = 0.20, CI [0.02, 0.36]. Overall, this suggests that it is more likely to observe more accurate judgments and decisions with increasing time on the task.  

\vspace{-0.5em}
\subsection{Participant Self-reported Strategies}
\label{sec:exps-strategiesUsed}
\color{black}

\revision{In Experiment 2, we found suggestive evidence for a difference between judgment and decision accuracy. To help explain our findings, we analyzed participants' self-reported strategies through a qualitative analysis. We identified high-level strategies and keywords related to judgment and decision making, including variants like \quotes{judge}, \quotes{decide}, and \quotes{choose}. A comprehensive breakdown of these strategies is available in the supplementary materials at the OSF link.}

\revision{Many participants in both judgment (61 of 146) and decision (62 of 155) tasks either did not answer or reported having no strategy. The most common strategy reported by both judgment (50 of 146) and decision (58 of 155) participants involved seeking for a balance between the cost/profit and probability. The textual analysis did not reveal differences between judgment and decision responses: 17 judgment participants mentioned decision-related terms, while 19 decision participants did so. It appeared that these references could simply relate to the AFC response choice format rather than being indicative of a high-level strategy.}

\vspace{-1em}
\subsection{Summary of Experiment 2}
Consistent with H${_1}$,  we found \revision{suggestive evidence} that, overall, decisions are more accurate than judgments. 
For H${_2}$ to H${_4}$, we noted in Sec.\,\ref{sec:exp1-resultsOfExperiment1} that, given the relatively large number of statistical hypotheses, 
each individual finding should be interpreted as \revision{tentative}. 
However, the results in the second column of Fig.\,\ref{fig:mean_accuracy} are remarkably consistent with our results for H${_1}$, with CIs being either inconclusive or providing evidence that \DA is more accurate. Thus there is converging evidence.
Importantly, we obtained different results for our conditions (\emph{e.g.}, the evidence is slightly stronger for the \condScenHumanitarian scenario than for the \condScenSports scenario), but we cannot conclude that the effect differs across 
conditions\,\cite{gelman2006difference,nieuwenhuis2011erroneous}.
Regarding H${_5}$, Fig.\,\ref{fig:mean_accuracy} suggests that the higher accuracy for the \DT might be driven by a stronger status quo bias in the \JT, indicating participants were more inclined to take action in the \DT. \revision{Finally, through exploratory qualitative analysis, we noted that participants did not appear aware of following a strategy that differentiates between judgment and decision.}

\vspace{-1em}
\section{Discussion}
\label{sec:discussion}

Experiment 1 did not provide evidence of a difference between \JA and \DA (Sec.\,\ref{sec:exp1-resultsOfExperiment1}). Experiment 2, with a slightly different design (including increasing the task difficulty, see Sec.\,\ref{sec:exp2}),  gave supportive evidence suggesting that judgment accuracy is not a good proxy for  decision accuracy. We discuss these findings next.

First, \textbf{\revision{we found suggestive evidence that} \DA was higher than \JA}. 
This outcome contrasts with Dimara \emph{et al.}'s findings\,\cite{Dimara2017a} 
(see Sec.\,\ref{sec:related_work_conflation}). In their study, both narratives involved an optimal selection  among 20 options. Participants in the decision making narrative, asked to  choose a house for themselves,   were less accurate than participants in the analytic narrative,  asked to identify a top real estate deal for clients.
Three factors might explain this discrepancy: 
(1) different numbers of alternatives (5 in our work \emph{vs.} 20 in\,\cite{Dimara2017a}); 
(2) differences between the nature of the scenarios (\condScenHumanitarian and \condScenSports \emph{vs.} real estate); and 
(3) different task complexities. Previous studies showed that the \DT gets harder with more alternatives, \emph{e.g.}, 10 or 20 in\,\cite{dy2021improving}. Secondly, a scenario's emotional value can affect how people approach a \DT (see the \condScenHumanitarian and \condScenSports scenarios). Finally, on task complexity, it appears that the \JT(s) in Dimara \emph{et al.} are not too hard, as they only require spatial comparisons of different points. In contrast, the \JT in our work require mental aggregation of extracted probabilities from uncertainty visualizations with cost/profit values to make a rough trade-off analysis -- which is less straightforward and prone to more errors. 

Secondly, \textbf{in the \condScenHumanitarian scenario, we found \minorrevision{suggestive} evidence that \DA was higher than \JA }. If a scenario effect indeed exists, it might stem from the emotional value associated with the \condScenHumanitarian scenario. Previously, Castro \emph{et al.}\,\cite{Castro} analyzed decision strategies and showed that emotions associated with alpacas in danger made some subjects act irrationally, \emph{e.g.}, by constantly issuing blankets to alpacas. Contrary to this, our findings suggest that emotions might not always be detrimental (see also\,\cite{lufityanto2016measuring}). Feeling responsible about deciding to help orphan kids might have sharpened participants' focus on pertinent information, aiding subsequent mental calculations. For example, some participants reported that they exclusively calculated EVs for the \condScenHumanitarian scenario: \quotes{\textit{I just tried to balance the profits/help for the kids with a probability of winning/the goods arriving on time}.}

Thirdly, \textbf{a stronger status-quo bias was \revision{potentially} more prevalent in the \JT than in the \DT}. 
While status-quo bias has been identified as a cognitive bias for decision tasks\,\cite{dimara2018task}, our results indicate its presence also in judgment tasks.
Research suggests that individuals tend to favor the status-quo especially when faced with challenging tasks\,\cite{fleming2010overcoming}. 
The strong bias in our \JT might stem from its inherent difficulty, given it demands mentally combining probabilities with cost/benefit evaluation. Conversely, the diminished status-quo bias in the \DT might be attributed to participants feeling a direct accountability for outcomes, driving them to take immediate actions. 

\revision{Among the three visualizations, \textbf{we found suggestive evidence for a difference only with probability bars, indicating that \DA was higher than \JA}.  This finding is consistent with Kale et al.\,\cite{Kale2021} study in that interval visualizations prompted better decisions.}
  
Furthermore, \textbf{we clearly observed that experiment 2 was harder than experiment 1}, as can be traced from the \revision{distribution of raw accuracy scores (see Fig.\,\ref{fig:raw_accuracy})}, even though our goal was not to compare the two experiments. This confirms that the changes we made, \emph{e.g.}, unique trials, and personalized task wording increased task difficulty. However, although the two experiments differed in their outcomes (inconclusive \emph{vs.} \DA $>$ \JA), there was no evidence of differences in effects between experiments. Therefore, we cannot conclude that the changes we made to experiment 1 amplified the difference between the \DT and the \JT\,\cite{nieuwenhuis2011erroneous,gelman2006difference}.

\revision{Additionally, designing an experiment to equitably compare judgment and decision making tasks was intricate. To enable such comparison, we simplified these tasks based on key decision making features discussed in previous research (see Sec.\,\ref{sec:jdmdefinitions}). Although we observed distinct trends in accuracy differences, this simplified design, which might explain why participants' self-reported strategies were not as informative (see Sec.\,\ref{sec:exps-strategiesUsed}), cannot reveal the underlying reasons for those differences. We encourage future research to explore these decision features and use more elaborated tasks to uncover the heuristics that decision makers apply.}

Nevertheless, our findings \revision{suggest} that even with simple judgment and decision tasks—those requiring only basic calculations on a few attributes—\DA appears higher than \JA, while the Dimara et al.\,\cite{Dimara2017a} study suggested the reverse.
Yet our findings align with both Kale \emph{et al.}\,\cite{Kale2021} and Dimara \emph{et al.}\,\cite{Dimara2017a} studies on the need to decouple judgment and decision making performance. 
We thus propose four essential action points for visualization research:
(1) establishing a shared \emph{definition} of what constitutes judgment and what decision making; 
(2) investigating \emph{task context} familiarity and valence to counteract potential biases; 
(3) developing \emph{scales} to measure and categorize difficulty levels for judgment tasks and  decision tasks, facilitating comparisons across studies and enabling the calibration of task difficulty; and 
(4) developing \emph{metrics} that assess a complex system's capacity to aid decision making. 
This diverges from current evaluations which use analytic task performance as a proxy for decision making ability or solely measure task-time.
For the latter, one could adapt preference elicitation methodologies\,\cite{Dimara2018DecisionSupport} to objectively measure users' \DA by their subjective preferences.

\vspace{-1.2em}
\section{Conclusion}

The conjoined study of judgment and decision making has a long history, much older than the visualization field\,\cite{kahneman1991article,kahneman2003perspective,gigerenzer2011heuristic}. Our work contributed to the decoupling of judgment and decision making in the following ways. We identified  inconsistencies in terminology and consequently   misinterpretations 
on whether people make suboptimal decisions because of or despite inaccurate judgments of information. To enhance clarity,  we analyzed, compared, and contrasted relevant concepts, experiment designs,  and findings from the literature. To the best of our knowledge, we  conducted  the first  experiments that investigated judgments and decisions  explicitly as distinct, yet direct-equivalent tasks.  Contrary to our expectations and trends observed  in previous research, we found decisions to be more accurate in affective  scenarios and  less vulnerable to the status-quo bias, suggesting that decision makers may disfavor responses associated with inaction. We conclude that \revision{judgments cannot be safely used as proxy tasks for decision making}.

Yet studying how, when, where, and why decision and judgment tasks differ remains an open question. Future research is needed to understand the currently unknown heuristics that decision makers use.

\bibliographystyle{ieeetr}
\bibliography{references.bib}

\vspace*{-0.2cm}
\begin{IEEEbiography}
[\vspace*{-1.2cm}
{\includegraphics
[width=1in,height=1in,clip,
keepaspectratio]{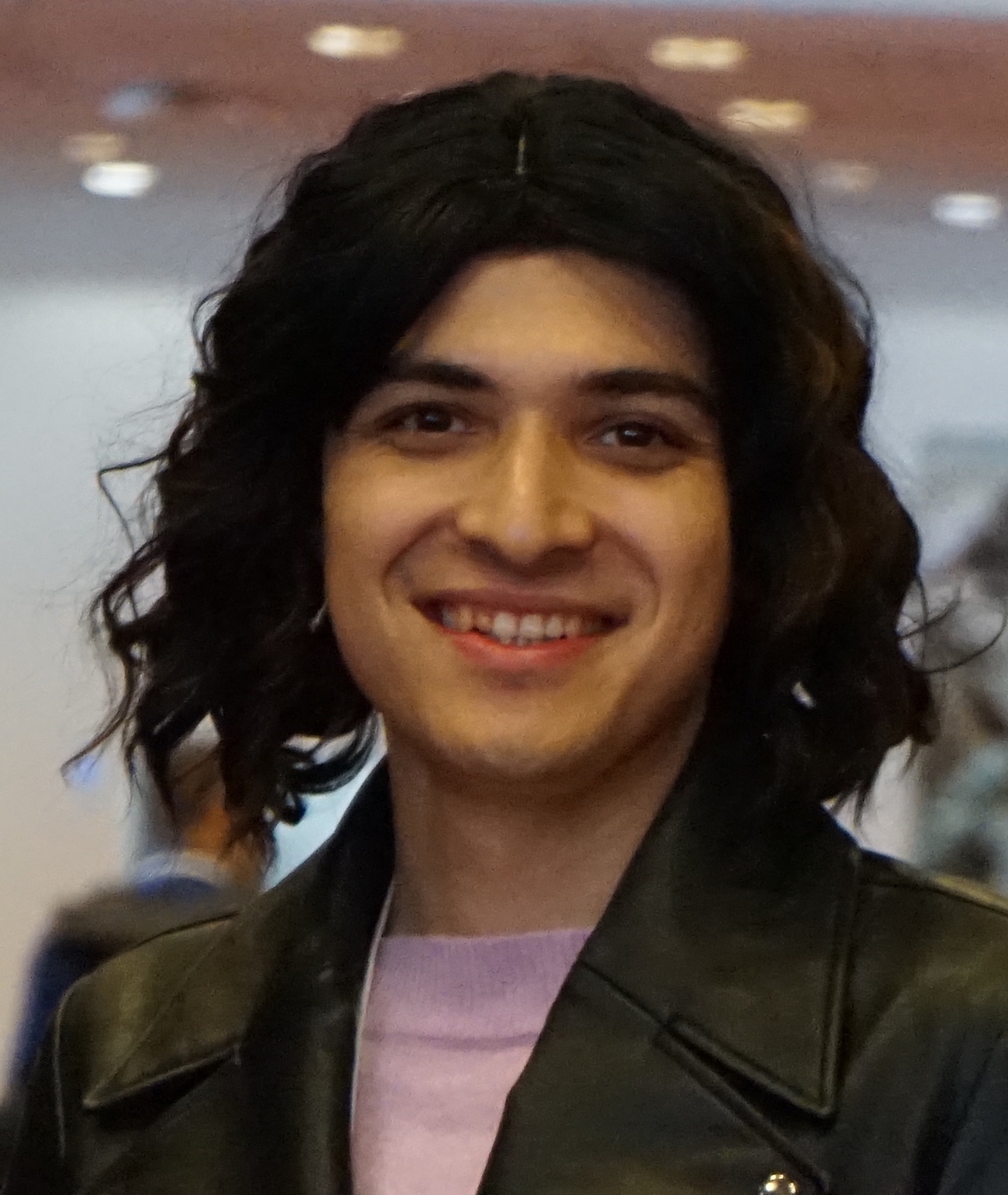}}]
{Başak Oral}
earned an MA in Cognitive Psychology from Bogazici University, Turkey, in 2021 and, since November 2021, has been enrolled at Utrecht University as a PhD candidate with a primary research focus on multi-criteria decision making (MCDM) with interactive data visualizations.
\end{IEEEbiography}

\vspace*{-1.8cm}
\begin{IEEEbiography}
[\vspace*{-0.5cm}
{\includegraphics
[width=1in,height=1in,clip,
keepaspectratio]{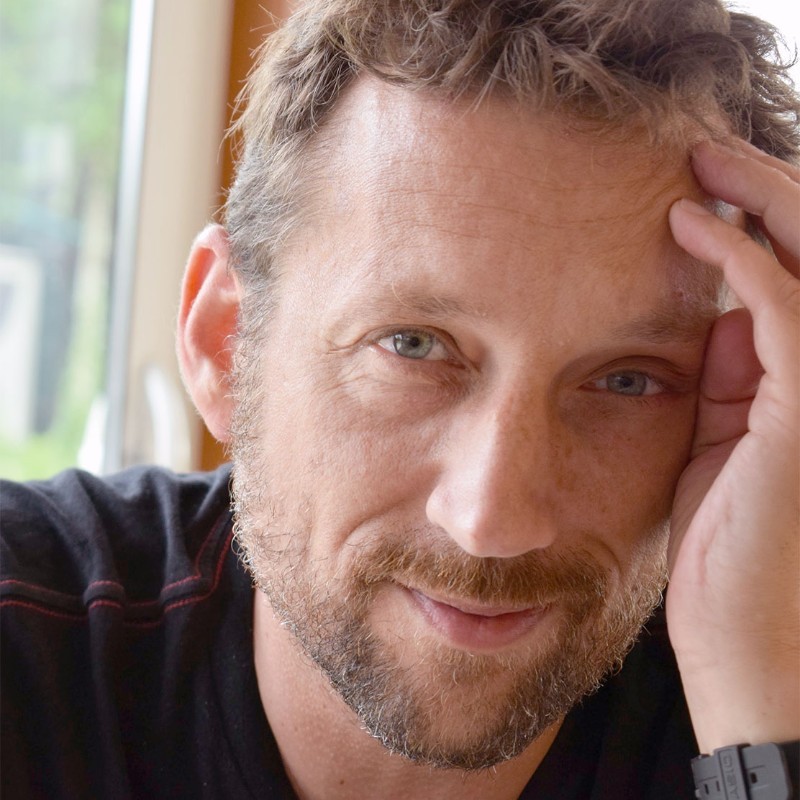}}]
{Pierre Dragicevic}
is a permanent Research Scientist at Inria Bordeaux, France. He is interested in humanitarian data visualization, physical and immersive visualizations, judgment and decision-making with visualizations, research transparency and statistical communication, as well as design spaces and conceptual frameworks.
\end{IEEEbiography}

\vspace{-1.6cm}
\begin{IEEEbiography}
[\vspace*{-0.1cm}
{\includegraphics
[width=1in,height=1in,clip,
keepaspectratio]{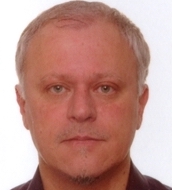}}]
{Alexandru Telea}
received his PhD (2000)
in Computer Science from the Eindhoven University of Technology. He was
assistant professor in visualization and computer
graphics at the same university (until 2007) and
then full professor of visualization at the University of Groningen. Since 2019 he is full professor of visual data analytics at
Utrecht University. His interests include high-dimensional visualization, visual analytics, and image-based information visualization. 
\end{IEEEbiography}

\vspace{-1.3cm}
\begin{IEEEbiography}
[\vspace*{-0.8cm}
{\includegraphics
[width=1in,height=1in,clip,
keepaspectratio]{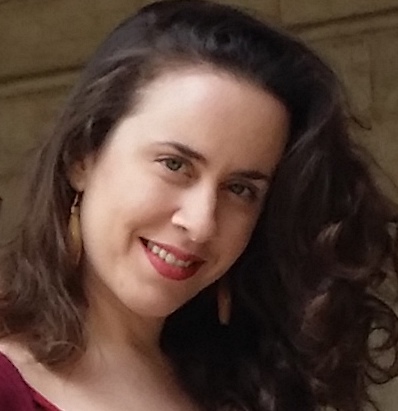}}]
{Evanthia Dimara}
is Tenured Assistant Professor at Utrecht University. Her fields of research are Information Visualization and Human-Computer Interaction. Her focus is on decision making -- how to help people make unbiased and informed decisions alone or in groups. 
\end{IEEEbiography}

\IEEEdisplaynontitleabstractindextext

\IEEEpeerreviewmaketitle

\vfill

\end{document}